# Ferroelectric-HfO$_2$/Oxide Interfaces, Oxygen Distribution Effect and Implications for Device Performance

**Notes: we have identified a mistake in the original version of manuscript entitled "Unexpected enhancement of ferroelectricity in HfO$_2$ on SiO$_2$ and GeO$_2$", which will directly compromise the outcome and affect the conclusion presented in the original manuscript. To be specific, we noticed that this model had not been built with symmetric oxygen distribution at the two interfaces in the supercell structure, but with two more oxygen ions at the Hf-interface than at the O-interface. This asymmetric oxygen distribution directly translates to additional net dipole across fe-HfO2 in the direction that opposes depolarizing field. Because this built-in field is intense, the overall field across fe-HfO2 tends to enhance its polarization. After revising this model to the one having symmetric oxygen distribution, depolarizing field emerges in fe-HfO2, similar to the cases in other fe-HfO2/oxide systems. Therefore, the originally claimed unexpected enhancement of HfO2 ferroelectricity is an artefact. As will be discussed in the revised manuscript, however, asymmetric oxygen distribution is highly relevant for fe-HfO2-based devices.**

Shihui Zhao[1+], Bowen Li[2+], Yuzheng Guo[3], Huanglong Li*[2,4]

[1] School of Materials Science and Engineering, Tsinghua University, Beijing, 100084, China [2] Department of Precision Instrument, Center for Brain Inspired Computing Research, Tsinghua University, Beijing, 100084, China [3] College of Engineering, Swansea University, SA1 8EN, Swansea, UK [4] Chinese Institute for Brain Research, Beijing, 102206, China

[+]equal contributions

*Email: li_huanglong@mail.tsinghua.edu.cn

**Abstract**

Atomic-scale understanding of $HfO_2$ ferroelectricity is important to help address many challenges in developing reliable and high-performance ferroelectric $HfO_2$ (fe-$HfO_2$) based devices. Though investigated from different angles, a factor that is real device-relevant and clearly deserves more attention has largely been overlooked by previous research, namely, the fe-$HfO_2$/dielectric interface. Here, we investigate the electronic structures of several typical interfaces formed between ultrathin fe-$HfO_2$ and oxide dielectrics in the sub-3-nm region. We find that interface formation introduces strong depolarizing fields in fe-$HfO_2$, which is detrimental for ferroelectric polarization but can be a merit if tamed for tunneling devices, as recently demonstrated. Asymmetric oxygen distribution-induced polarity, intertwined with ferroelectric polarization or not, is also investigated as a relevant interfacial effect in real device. Though considered detrimental from certain aspects, such as inducing build-in field (independent of ferroelectric polarization) and exacerbating depolarization (intertwined with ferroelectric polarization), it can be partly balanced out by other effects, such as annealing (extrinsic) and polarity-induced defect formation (intrinsic). This work provides insights into ferroelectric-$HfO_2$/dielectric interfaces and some useful implications for the development of devices.

The replacement of $SiO_2$ with high-K $HfO_2$ as the gate insulator in 2007 has been considered as one of the biggest redesign of the transistor technology[1], enabling continuous down-scaling of the transistor according to the prediction of Gordon Moore[2]. Today, aggressive transistor scaling down to 3nm-scale is on the way[3] at which the semiconductor community is becoming increasingly concerned about the ending of Moore's law. Reinvention of the device technology is imperatively called for[4].

In 2011, a group of German researchers discovered in Si-doped $HfO_2$ unexpected ferroelectricity[5, 6], a rare materials property that has previously only been found in complex crystals, such as perovskites, which are unfortunately not CMOS-compatible. Because $HfO_2$ had already been a key enabler of the high-K-metal-gate technology for transistors, this discovery has led to a re-emergence of intensive research interest in ferroelectric devices, including negative-capacitance (NC) transistor[7-10], ferroelectric memory transistor (or fe-transistor)[11-14] and ferroelectric tunneling junction (FTJ) nonvolatile memory[15, 16].

Although it is the intrinsic ferroelectricity in $HfO_2$ that underpins the workings of these devices, ferroelectric $HfO_2$ (fe-$HfO_2$) cannot be considered in isolation, but rather the characterized device performance reflects the entire system of the component films and the interfaces between them[17]. In the past, extensive investigations on ferroelectric/electrode interfaces have been conducted. Ferroelectric/electrode interfaces influence device performance in many key aspects, including critical film thickness (power and scalability relevant)[18], field-cycling behavior[19], tunneling electroresistance (TER, for FTJ)[20], and so on. In addition to ferroelectric/electrode interfaces, it is not hard to find that fe-$HfO_2$ is most often used in combination with another non-ferroelectric dielectric in a real device for practical reasons, giving rise to fe-$HfO_2$/dielectric interface. For example, NC-transistor typically uses series fe-$HfO_2$/dielectric gate capacitor[21-25] to provide sufficient capacitance matching and maintain the overall gate capacitance positive[7-10], and fe-transistor typically uses fe-$HfO_2$/dielectric double-layer, in particular, fe-$HfO_2$/$SiO_2$[26-30], for improved quality of the gate stack/semiconductor interface[31], and in FTJ fe-$HfO_2$/dielectric double-layer[32] can be used to achieve enhanced TER[33].

More broadly, dielectric substrates are very common for the growth of ferroelectric thin films. They interact intimately, so much so that this has usually been taken for granted. In a recent report on the counterintuitive enhancement of ferroelectricity in 1-nm ultrathin Zr-doped $HfO_2$ grown on $SiO_2$/Si[34], the formation of the fe-$HfO_2$/$SiO_2$ interface has only been mentioned marginally.

Furthermore, ferroelectric/dielectric interfaces are the places where electronic charge degree of freedom is strongly coupled to ionic lattice degree of freedom, creating various novel emergent phenomena, such as the formation of two-dimensional electron/hole gas (2DEG/2DHG)[35-37]. In this respect, ferroelectric/dielectric interface has a well-known analogue, i.e., the dielectric/dielectric interface across which polar discontinuity exists, such as $LaAlO_3$/$SrTiO_3$ interface[38]. The difference is that the interfacial property in the ferroelectric/dielectric system is electrically switchable via ferroelectric switching. This enhances functional flexibility, and of course, potentially introduces more intricacies. Recently, ferroelectric switching in fe-$HfO_2$ has been unambiguously shown to be intertwined with electrochemically driven oxygen migration[39]. This may either enhance or alleviate polar discontinuity, dependent on whether asymmetric oxygen distribution-induced net dipole across fe-$HfO_2$ has the same direction as fe-$HfO_2$ polarization.

In this work, we investigate several representative fe-$HfO_2$/oxide dielectric interfaces by first-principles calculations. The oxide dielectrics under consideration cover a range of cation valence, electronegativity and band gaps. As a relevant interfacial effect in real device, asymmetric oxygen distribution-induced polarity, intertwined with ferroelectric polarization or not, is also investigated. With their electronic structures calculated, we discuss on their potential implications for fe-$HfO_2$-based devices.

We create supercell models of several fe-$HfO_2$/oxide interfaces, including fe-$HfO_2$/$Al_2O_3$, fe-$HfO_2$/$La_2O_3$, fe-$HfO_2$/$SiO_2$ (quartz), fe-$HfO_2$/$GeO_2$, fe-$HfO_2$/m-$HfO_2$ (m for monoclinic), fe-$HfO_2$/t-$HfO_2$ (t for tetragonal), fe-$HfO_2$/$TiO_2$, fe-$HfO_2$/$P_2O_5$, fe-$HfO_2$/$Ta_2O_5$ and fe-$HfO_2$/$WO_3$, each containing roughly 25 Å thick fe-$HfO_2$ and oxide with comparable thickness. Note that this modelled thickness value should not be viewed as being too small because fe-$HfO_2$ has already been scaled down to the sub-3-nm region[34, 40] and FTJ device based on 1-nm fe-$HfO_2$/1-nm $SiO_2$ has also been demonstrated[32]. The fe-$HfO_2$ model is selected to be in the polar orthorhombic $Pca2_1$ phase which has generally been considered as the source of ferroelectricity[41, 42], while other ferroelectric phases have also been proposed[43, 44]. In this fe-$HfO_2$ model, the displacements of the three-fold coordinated oxygen ions ($O_{III}^{2-}$) along the c axis give rise to ferroelectric polarization. The electron-counting-rule is satisfied in each model. The crystalline phases of the non-ferroelectric oxides used for building the corresponding interface models are listed in supplementary table S1. The oxide slabs are cleaved along the surface orientations (listed in supplementary table S1) that provide interfacial bonding environments as compatible as possible and achieve reasonable tradeoff between lattice matching and computational burden. The in-plane lattice constants are fixed to that calculated for fe-$HfO_2$ single crystal, and the lattice perpendicular to the interface is relaxed, using the CASTEP code[45]. The calculations use the Perdew-Burke-Ernzerhof (PBE) flavor of the generalized gradient approximation (GGA)[46], N×M×1 Monkhorst-Pack k-points with in-plane spacing smaller than 0.067 1/Å, and ultrasoft pseudopotentials with a plane wave cutoff energy of 380 eV. Convergence test validates the parameterization, as shown in the supplementary figure S1.

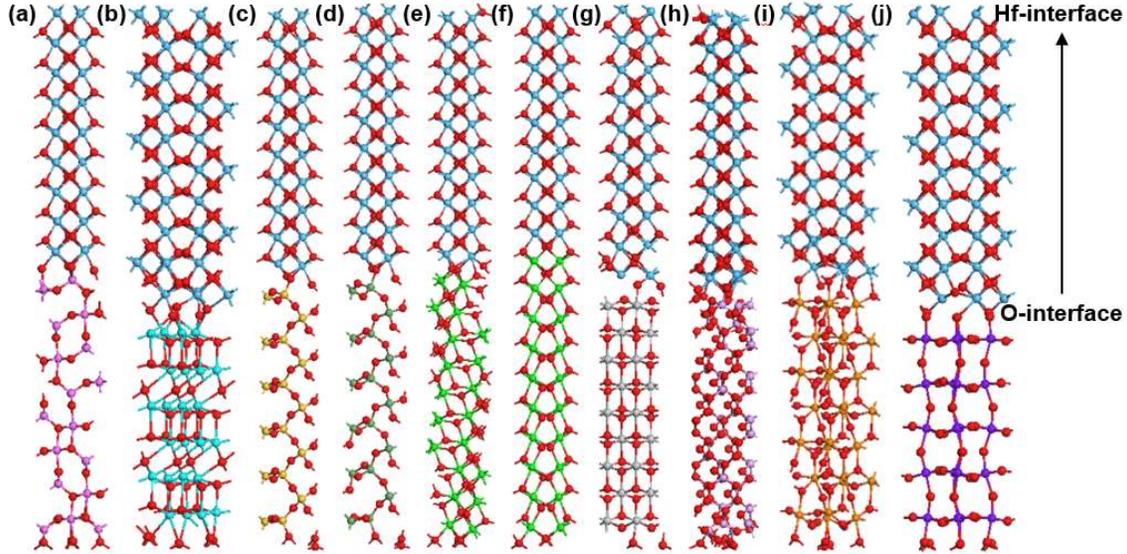

**Figure 1** Atomic structures of the (a) fe-$HfO_2$/$Al_2O_3$, (b) fe-$HfO_2$/$La_2O_3$, (c) fe-$HfO_2$/$SiO_2$, (d) fe-$HfO_2$/$GeO_2$, (e) fe-$HfO_2$/m-$HfO_2$, (f) fe-$HfO_2$/t-$HfO_2$, (g) fe-$HfO_2$/$TiO_2$, (h) fe-$HfO_2$/$P_2O_5$, (i) fe-$HfO_2$/$Ta_2O_5$ and (j) fe-$HfO_2$/$WO_3$ interface systems. The arrow shows the direction of ferroelectric polarization.

In all interface models, the polarizations of fe-$HfO_2$ are sustained after geometry optimizations, as seen in figure 1(a-j). Each supercell model contains two interfaces that look symmetric in terms of atomic packing but actually asymmetric because of being polar. All these models are shown with the polarizations of fe-$HfO_2$ in the upward directions. The two asymmetric interfaces in each model with $Hf^{4+}$ bound charge and $O_{III}^{2-}$ bound charge are referred to as the Hf-interface and O-interface, respectively. For the sake of convenience, the interface is also referred to as the end interface or the middle interface according to its position in the supercell. In models shown in figure 1, the O-interfaces are the middle interfaces and the Hf-interfaces are the end interfaces. The projected density of states (PDOS) of each model is shown in figure 2(a-j). For fe-$HfO_2$/m-$HfO_2$, the bands of fe-$HfO_2$ bend toward higher energy as they approach the O-interface. Similarly, the bands of m-$HfO_2$ also bend toward higher energy as they approach the O-interface. These result in wedge-shaped potential wells for electrons and holes at the Hf-interface and O-interface, respectively. The Fermi energy ($E_f$) of this interface system lies close to the conduction band minimums (CBMs) of fe-$HfO_2$ and m-$HfO_2$ at the Hf-interface, and lies close to their valence band maximums (VBMs) at the O-interface, indicating the tendency to form 2DEG and 2DHG, respectively. Experimentally, 2DEG and 2DHG have indeed been observed at the perovskite ferroelectric/dielectric interface system[35, 37].

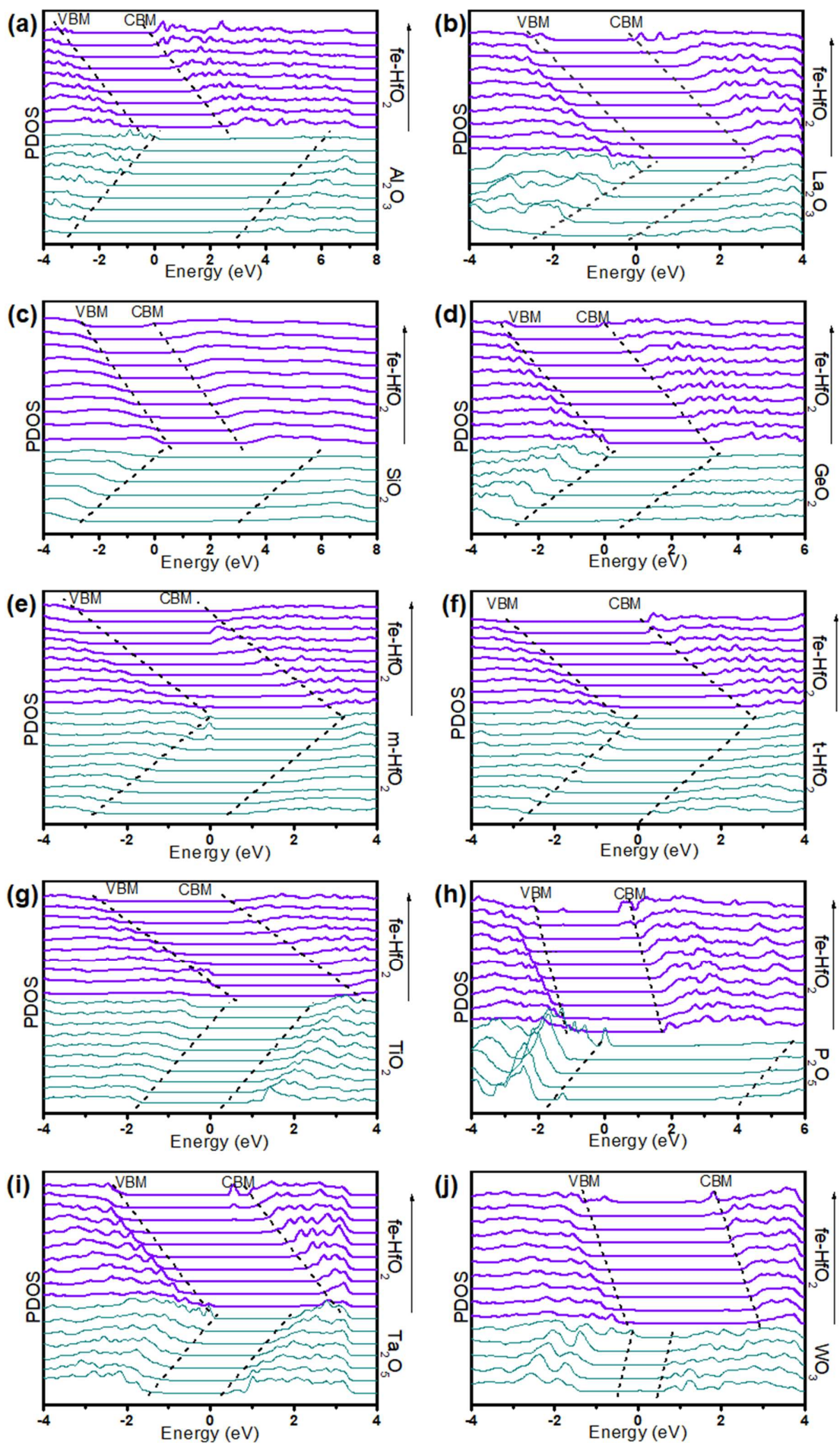

**Figure 2** PDOSs of the (a) fe-HfO$_2$/Al$_2$O$_3$, (b) fe-HfO$_2$/La$_2$O$_3$, (c) fe-HfO$_2$/SiO$_2$, (d) fe-HfO$_2$/GeO$_2$, (e) fe-HfO$_2$/m-HfO$_2$, (f) fe-HfO$_2$/t-HfO$_2$, (g) fe-HfO$_2$/TiO$_2$, (h) fe-HfO$_2$/P$_2$O$_5$, (i) fe-HfO$_2$/Ta$_2$O$_5$ and (j) fe-HfO$_2$/WO$_3$ interface systems, respectively. The Fermi energies are set to 0 eV.

The direction of band bending in fe-HfO$_2$ indicates the existence of depolarizing field that tends to destabilize its ferroelectric polarization. In the case of fe-HfO$_2$/m-HfO$_2$, 2.2-V voltage drop across 2.7-nm fe-HfO$_2$ will translate to 8.3-MV/cm depolarizing field which is larger than the coercive field measured experimentally[47]. Interface formation induces depolarizing field due to incomplete screening of the bound charge[48, 49]. The tendency to form 2DEG (2DHG) at the Hf-interface (O-interface) in ferroelectric/dielectric stack can be understood as a way to screen the bound charge.

Electric field in opposite direction to the depolarizing field in fe-HfO$_2$ is also built up in m-HfO$_2$. This field is also sizable, reaching 7.8 MV/cm. In some cases (see supplementary table S2 for field intensities in various fe-HfO$_2$/oxide interfaces), the fields in dielectrics are so strong that they could induce dielectric breakdown. It has been proposed that in practice the field in dielectric will maintain below the breakdown value because free carrier leakage through the dielectric and the ensuing trapping at the ferroelectric/dielectric interface can provide additional screening of the bound charge[31].

Upon fe-HfO$_2$ ferroelectric switching, both the depolarizing field in fe-HfO$_2$ and the field in m-HfO$_2$ will change directions. It is the reversal of field and the ensuing large potential difference experienced by the m-HfO$_2$ layer that give rise to large TER in ferroelectric/dielectric-based FTJ device[33].

The fe-HfO$_2$/t-HfO$_2$ interface system shows similar band structure properties to those of the fe-HfO$_2$/m-HfO$_2$ system. The dynamic structural transition from m-HfO$_2$ or t-HfO$_2$ to fe-HfO$_2$ has been considered to be vital in giving rise to the field-cycling behavior of fe-HfO$_2$-based ferroelectric capacitors[19]. In particular, it increases the ferroelectric switching response during the wake-up phase of the device. This is easy to understand because m-HfO$_2$ and t-HfO$_2$ are not ferroelectrically active. From an interface perspective, our results also indicate that the pre-existing fe-HfO$_2$/m-HfO$_2$ and fe-HfO$_2$/t-HfO$_2$ grain boundaries will degrade the ferroelectricity in fe-HfO$_2$, but with gradual structural transition to fe-HfO$_2$ upon cycling this effect will be mitigated due to the decrease of depolarizing field.

For interfaces between fe-HfO$_2$ and other metal-oxide dielectrics, namely, Al$_2$O$_3$, La$_2$O$_3$, SiO$_2$, GeO$_2$, TiO$_2$, Ta$_2$O$_5$ and WO$_3$, similar interfacial band structure properties to those of fe-HfO$_2$/m-HfO$_2$ and fe-HfO$_2$/t-HfO$_2$ interfaces are found.

In periodic supercell modelling approach, a two-terminal device structure with short-circuited electrodes can be naturally modeled[18]. Our periodic supercell model is analogous to such a fe-HfO$_2$/oxide-based device whose electrodes are ideal (zero Thomas–Fermi screening length) and infinitely thin (electrodes vanished so that fe-HfO$_2$ and oxide are brought together to form the end interface, due to periodicity). As introduced, each of the above supercell models is built to have the two interfaces with similar atomic packing. In doing so, the two interfaces are assumed to have equal atomic chemical potentials. However, this might not always be the case. In fact, a relatively high and low oxygen chemical potentials (μ$_O$) could exist at the fe-HfO$_2$/oxide interface (middle interface), and at the fe-HfO$_2$/electrode and oxide/electrode interfaces, respectively, in an as-fabricated device.

We simulate this condition in fe-HfO$_2$/SiO$_2$ system (the arguments developed here are very general and could be readily extrapolated to other fe-HfO$_2$/oxide systems) by making the middle interface O-rich and the end interface O-poor while keeping the electron-counting-rule satisfied. The difference in the number of O ions between these two interfaces is two. Bulk quartz SiO$_2$ is a three-dimensionally connected network with alternating short and long Si-O bonds[50]. Its basic structural motif is the irregular SiO$_4$ tetrahedron with shared corners. GeO$_2$ is a network glass with similar properties to SiO$_2$. Compared to GeO$_2$, SiO$_2$ has less irregular SiO$_4$ tetrahedrons[51], as seen from figure 1(c, d). In the case of fe-HfO$_2$ polarization pointing away from SiO$_2$ (up-polarization), as shown in figure 3a, depolarizing field is built up in fe-HfO$_2$ (figure 3b), reaching 11.8 MV/cm, which is stronger than 10.8 MV/cm in the case of equal interfacial μ$_O$. Such large depolarizing field indicates that fe-HfO$_2$ up-polarization is highly unfavorable. However, in the case of fe-HfO$_2$ down-polarization, as shown in figure 3c, hyperpolarizing (anti-depolarizing) field is built up in fe-HfO$_2$ (figure 3d), tending to enhance fe-HfO$_2$ down-polarization. Therefore, under this fabrication-relevant μ$_O$ condition, the fe-HfO$_2$/SiO$_2$ system is mono-stable in zero applied field that only one of the two polarization states is stable. This mono-stability originates from the asymmetric μ$_O$ at the two interfaces (forming a net dipole) and the consequential built-in field in fe-HfO$_2$, reminiscent of the cases in other non-ferroelectric systems[38, 52-54]. The built-in fields in our models are strong because the μ$_O$ difference between the two interfaces is large, one O per surface Hf atom. In practice, μ$_O$ difference may not be that large, and may not survive annealing considering that HfO$_2$ is a fast oxygen ion conductor at high temperature. Experimentally, annealing is conveniently used to crystalize fe-HfO$_2$, and, according to our simulations, to level out μ$_O$ difference and decrease the built-in field. Nevertheless, for cases in which high-temperature processes are undesired, such as flexible electronics on organic substrates, the influence of this μ$_O$ difference can be significant.

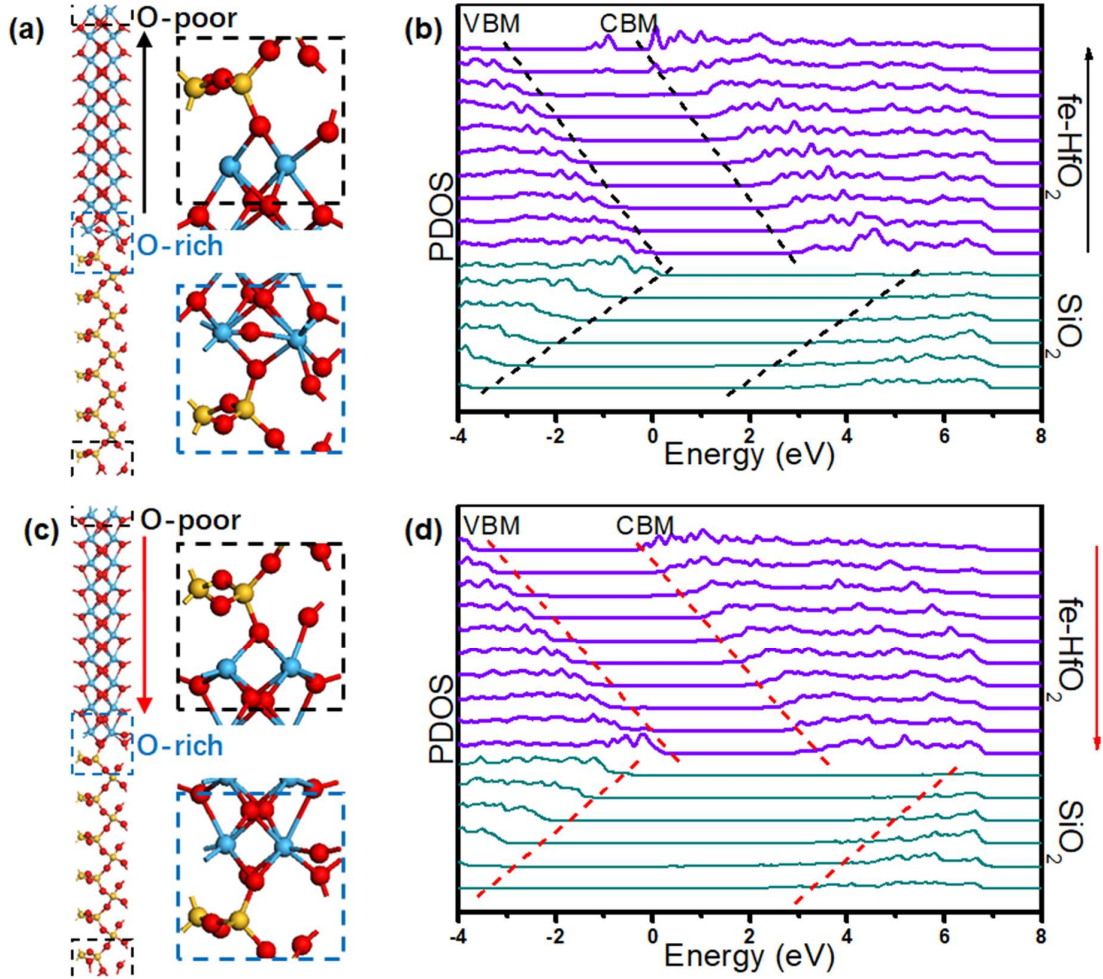

Figure 3 Atomic structure of the fe-HfO$_2$/SiO$_2$ system with O-rich middle interface and O-poor end interfaces, the fe-HfO$_2$ polarization is (a) upward and (c) downward. PDOSs corresponding to (b) up-polarization and (d) down-polarization configurations.

This $\mu_O$ difference can even change dynamically with the ferroelectric switching process. It has been known that ferroelectric switching is intertwined with oxygen vacancy migration[55] and this migration can be very directional subject to electrochemical forces[39, 56]. In the presence of such intertwined and directional oxygen vacancy migration, the O-interface will always be O-rich and the Hf-interface will always be O-poor. This is a similar case to that shown in figure 3a. The difference is that the asymmetric $\mu_O$ distribution-induced net dipole across fe-HfO$_2$ changes reversibly with ferroelectric switching. Thus, both fe-HfO$_2$ polarization states are destabilized. This seems to impose a fundamental challenge for fe-HfO$_2$ devices. However, there is a compensating effect. As mentioned earlier, in the ideal fe-HfO$_2$/oxide system the O-interface and Hf-interface are relatively p-type and n-type, respectively, due to polarity-induced band bending (figure 1). It is known that the formation energies and charge states of point defects are strongly dependent on the electronic chemical potential (Fermi energy). Alam et al. have shown that in fe-HfO$_2$ positively charged oxygen vacancies and negatively charged oxygen interstitials are thermodynamically easy to form under p-type and n-type conditions, respectively[57]. These are exactly the conditions provided by the O-interface and Hf-interface in the ideal fe-HfO$_2$/oxide system, respectively. Therefore, the thermodynamically triggered formation of the corresponding defects at these two interfaces will partly balance out the electrochemical migration-induced difference in $\mu_O$. Note that polarity-induced formation of oxygen vacancy defects has been considered as a plausible mechanism in regulating the interfacial electronic structure in other non-ferroelectric systems[58, 59].

To conclude, we investigate several representative fe-HfO$_2$/oxide interfaces and find in all systems the existence of strong depolarizing field in fe-HfO$_2$ due to incomplete screening of the bound charge. The depolarizing field is detrimental that it destabilizes ferroelectric state, especially in the ultrathin region. Attention should be paid to this problem in developing NC-transistors and fe-transistors that uses fe-HfO$_2$/oxide bilayer structures. In an as-fabricated fe-HfO$_2$ device, the depolarizing field in fe-HfO$_2$ grain due to the formation of fe-HfO$_2$/m-HfO$_2$ and fe-HfO$_2$/t-HfO$_2$ grain boundaries is also an important factor influencing the device cycling behavior. Along with the depolarizing field, strong field in opposite direction is built up in oxide layer. For FTJ, the switchable electric potential across the oxide layer upon ferroelectric switching can be exploited as a source for TER[33], as long as ferroelectric polarization is not completely depolarized. As a relevant interfacial effect in real device, asymmetric oxygen distribution-induced polarity, intertwined with ferroelectric polarization or not, is also investigated. In an as-fabricated device, graded oxygen chemical potential due to different fabrication conditions results in built-in field that will make the system only mono-stable under zero applied field. This effect can be conveniently mitigated by annealing considering that HfO$_2$ is a good oxygen ion conductor. Related to its oxygen ion conducting property, ferroelectric switching in fe-HfO$_2$ is intertwined with reversible oxygen migration[39] that can dynamically change oxygen distribution. Despite its hidden benefits for novel electronic device functions, oxygen migration-induced net dipole across fe-HfO$_2$ can be troublesome in the sense of exacerbating the depolarization of fe-HfO$_2$. A polarity-induced interfacial defect formation mechanism is available to balance out these adversarial effects. This work provides insights into ferroelectric-HfO2/dielectric interfaces and some useful implications for the development of devices.

**Conflict of Interest**

The authors declare no competing financial interest.

**Acknowledgments**

The authors acknowledge funding from National Natural Science Foundation (grant no. 61974082, 61704096 and 61836004). The authors acknowledge funding from National Key R&D Program of China (2018YFE0200200), Youth Elite Scientist Sponsorship (YESS) Program of China Association for Science and Technology (CAST) (no. 2019QNRC001), supercomputing wales project number scw1070, Tsinghua-IDG/McGovern Brain-X program, Beijing science and technology program (grant no. Z181100001518006 and Z191100007519009), the Suzhou-Tsinghua innovation leading program 2016SZ0102, and CETC Haikang Group-Brain Inspired Computing Joint Research Center.

## DATA AVAILABILITY

The data that support the findings of this study are available from the corresponding author upon reasonable request.